\documentclass[12pt]{article}
\usepackage{epsf}

\newcommand{\lsim}{\raisebox{-.5ex}{\footnotesize$
     \,\:\stackrel{\textstyle<}{\sim}\,\:$}}

\begin{document}

\begin{flushright}
NIKHEF--97--027
\vspace{-10pt}
\end{flushright}

\begin{center}

{\LARGE
Nonperturbative $\gamma^*p$ Interaction\\
in the Diffractive Regime\\}
\vskip 20pt
{\Large
H.G. Dosch$^1$, T. Gousset$^2$ and H.J. Pirner$^1$}\\
\bigskip 
{\it
1- Institut f\"ur Theoretische Physik der Universit\"at Heidelberg,\\
Philosophenweg 16 \& 19, 69120 Heidelberg, Germany\\
\medskip
2- NIKHEF, P.O. Box 41882, 1009 DB Amsterdam, The Netherlands}
\bigskip
\end{center}

\begin{abstract}
One of the challenging aspects of electroproduction at high-energy is
the understanding of the transition from real photons to virtual photons
in the GeV$^2$ region. We study inclusive electroproduction on the
proton at small $x_B$ using a nonperturbative dipole-proton cross
section calculated from the gauge invariant gluon field correlators as input. 
By quark-hadron duality, we construct a photon light cone wave function
which links the ``hadronic'' behavior at small $Q^2$ to the
``perturbative'' behavior at large $Q^2$. It contains quark
masses which implement the transition from constituent quarks at low
$Q^2$ to current quarks at high $Q^2$. Our calculation gives a good
description of the structure function at fixed energy for 
$Q^2\leq 10$~GeV$^2$. Indications for a chiral transition may already 
have been seen in the photon-proton cross section.
\end{abstract}

\section{Introduction}

Since the late 60's, multi-GeV electron and muon collisions with protons
have been intensively used in order to get information on the 
{\em proton structure}. For photon virtualities $Q$ far below the
$Z$-mass, the interaction is mediated by a virtual photon and the
inclusive photon-proton cross section can be described by means of the
proton structure functions. At large $Q^2$, the leading unpolarized 
structure function, $F_2(x_{\rm B},Q^2)$, is to leading-log accuracy the
well-known linear combination of partonic distributions, 
$q_f(x_{\rm B},Q^2)$, weighted by the square of the parton
electromagnetic charge expressed in units of the proton charge:
$$
F_2(x_{\rm B},Q^2)=\sum_f \hat{e}_f^2 x_{\rm B}q_f(x_{\rm B},Q^2).
$$
An illustrative partonic description emerges when the process is
envisaged in a frame where the proton has a large momentum $\bf p$
(formally $|{\bf p}|\to\infty$) and in a particular gauge. At small
$Q^2$, the above decomposition and the partonic interpretation of the
process get spoiled by power corrections.

There is, yet, an alternative to the infinite momentum frame description
of the collision which is a description in the center of mass frame. In
this frame, the photon acquires a structure and we have to deal with the
interaction of {\em two structured objects}. Although it may look as if
we had not gained anything by changing our point of view, the operation
is interesting if we focus on the high-energy fixed $Q^2$ kinematical
domain of the process. In this regime, the bulk of the photon-proton 
interaction ressembles that of hadron-hadron, i.e., {\em diffractive
scattering}. In the Regge approach which is applicable in the
kinematical region envisaged here, this is understood as being due to
the universality of the Pomeron. Implicitely, this assertion assumes
that the photon has developed an internal structure due to its coupling
to strongly interacting quark fields and that this structure gives the
main contribution at small $x_{\rm B}$ to its interaction~\cite{bau78},
as compared to the direct contribution from its ``bare component''.
Therefore a great deal of insight can be gained from a common
understanding of both photon-hadron and hadron-hadron collisions. 

In Ref.~\cite{dos94}, the application to diffractive scatterings of 
hadrons of the model of the stochastic vacuum has been carried out. 
Recently, in Ref.~\cite{dos97}, the same approach was used to describe
diffractive leptoproduction of vector mesons in the range
$Q^2=2$--10~GeV$^2$, thus starting to implement the program just
mentioned. Our aim in the present paper is twofold: we want to pursue
the comparison by considering the total photon-proton cross section and
we want to extend the phenomenology to $Q^2\to 0$.

In Ref.~\cite{dos97}, the interaction amplitude for the exclusive 
vector meson photoproduction off a proton has been written as
\begin{equation}\label{amplitude}
{\cal M}(\gamma^*+p\to V+p)=is\int {dzrdr\over 2}
\psi_V^*\psi_\gamma(z,r)\,J^{(0)}_p(z,r,t).
\end{equation}
$\psi_V$ and $\psi_\gamma$ are, respectively, the vector meson and
virtual photon light cone wave functions. If the final vector meson wave
function is replaced by the virtual photon wave function, one gets the
forward Compton amplitude, that is $is\sigma^{\rm tot}_{\gamma^*p}$. The
quantity $J^{(0)}_p(z,r,t)$ represents the Pomeron exchange
amplitude for scattering of a $q\bar{q}$ dipole of size $r$ off the
proton target, where an average over the dipole orientations has been
carried out. The light cone fraction carried by the quark in the photon
is denoted by $z$. The Mandelstam variables for the process are
$s=W^2$ and $t$. The photon is characterized by its virtuality $Q^2$ and
polarization. 

The quantity $J^{(0)}_p$ has been derived in the model of the stochastic
vacuum~\cite{dos97} following the method of Ref.~\cite{dos94}. In these
references, the few parameters which fix the magnitude and shape of
$J^{(0)}_p$ have been adjusted to fit the phenomenology of proton-proton
elastic cross section. We use the same parametrization here and focus
on the photon structure.

Let us discuss the photon wave function that enters in
Eq.~(\ref{amplitude}). There are two standard schemes. The first one is
to expand the photon wave function in a hadronic basis~\cite{bau78}. The
wave function for a transversely polarized photon reads
\begin{equation}\label{photon_had}
\psi^{\rm had}_{\gamma(T)}=
\sum_{\rho,\omega,\phi}{ef_VM_V\over M_V^2+Q^2}\,\psi_{V(T)}
+\hbox{\em Rest},
\end{equation}
where we have explicitly written the low-lying vector meson
contribution. The symbol {\em Rest} stands for a sum over residual
$1^-$~states, like higher radial and orbital excitations and nonresonant
multiparticle states. According to the vector meson dominance (VMD)
hypothesis, the low-lying vector meson states, $\rho$, $\omega$, $\phi$,
dominate the photon wave function at small $Q^2$ making
Eq.~(\ref{photon_had}) a useful expansion in this regime. The wave
function for a longitudinally polarized photon is obtained by changing
$ef_VM_V\to ef_VQ$, $\psi_{V(T)}\to\psi_{V(L)}$ and the {\em Rest} term
accordingly.

To assess the relative importance of the residual contribution
{\em Rest} at large $Q^2$, we examine first the experimental behavior
of the inclusive cross section and the ratio 
$R^{tot}=\sigma_L/\sigma_T$ in the range $Q^2=1$--10~GeV$^2$ (cf. left
part of Table~\ref{behavior}). At large $Q^2$, the success of the parton
model with spin $1/2$ quarks tells us that the structure function 
$F_2(x)\propto Q^2\sigma(Q^2)$ scales, i.e., the total cross section
decreases as $Q^{-2}$ and that it is dominated by transverse photon
scattering. In VMD the contributions of $\rho$, $\omega$, $\phi$ alone
would lead to a $\gamma^*$-$p$ total cross section 
$\sigma_T\propto Q^{-4}$ and $\sigma_L\propto Q^{-2}$, i.e., a
dominating longitudinal cross section. It means that in the VMD
description the transverse inclusive cross section must be built from
the residual term {\em Rest} in the photon wave function. Let us next
consider vector meson production (cf. right part of
Table~\ref{behavior}). In the range $Q^2=1$--10~GeV$^2$ the experimental
cross section has a $Q^{-4}$ behavior and is predominantly longitudinal.
This dependence is overshooted by the longitudinally dominated VMD cross
section $O(Q^{-2})$. It thus turns out that the {\em Rest} term in the
photon wave function Eq.~(\ref{photon_had}) is also needed to cancel the
$\rho,\omega,\phi$ contribution in order to provide the right $Q^{-4}$
behavior for the elastic vector meson production. This cancellation 
mechanism has to be implemented in a consistent way to get a unique 
description of the photon in high energy scattering.

\begin{table}[ht]
$$
\begin{tabular}{|l|cccc|}
\hline
   &$\sigma^{\rm tot}$&$R^{\rm tot}$&$\sigma^{\rm el}$&$R^{\rm el}$\\
\hline
exp&$Q^{-2}$&$\le 0.2$&$Q^{-4}$&$>1$ and rising\\
VMD&$Q^{-2}$&$Q^2$&$Q^{-2}$&$Q^2$\\
free $q\bar{q}$&$Q^{-2}$&const&$Q^{-6}$&$Q^2$\\
\hline
\end{tabular}
$$
\caption{\label{behavior}
{\small Large $Q^2$ behavior of the cross section
$\sigma=\sigma_L+\sigma_T$ and of the longitudinal to transverse ratio
$R=\sigma_L/\sigma_T$ for both total photon-proton cross section and
elastic vector meson production. The first line shows the experimental
results in the range $Q^2=1-10\,$GeV$^2$. The second line displays the
asymptotic behavior expected from VMD. For comparison, the third line
gives the scaling behavior determined with a free $q\bar{q}$ wave
function (up to logarithms).}}
\end{table}
  
A possible solution is to use a quark-gluon basis. At values of $Q^2$
large enough, the hadronic component of the photon is indeed a free
$q\bar{q}$ pair, $\psi^{\rm had}_{\gamma}\approx\psi_{q\bar{q}}$, so
that the quark-gluon basis is more efficient than the hadron basis and
the cancellation mechanism occurs automatically. It seems
phenomenologically possible to envisage a kind of hybrid description
where the photon state can be viewed as a superposition of a few
low-lying resonances plus a free $q\bar{q}$-state. We stress, however,
that identifying the {\em Rest} in Eq.~(\ref{photon_had}) with the free
$q\bar{q}$ wave function $\psi_{q\bar{q}}$ is not sufficient because, on
the one hand, it would not reproduce the phenomenology given in
Table~\ref{behavior}, and, on the other hand, it leads to a double
counting of some hadronic configurations. In order to avoid these
problems, one has schematically to modify Eq.~(\ref{photon_had}) in such
a way that, for small values of $Q^2$, it agrees with  the
vector-dominance like form
$$
\psi^{\rm had}_{\gamma}=
\sum_{\rho,\omega,\phi}{ef_VM_V\over M_V^2+Q^2}\,\psi_{V},
$$ 
and, for large $Q^2$, it approaches the perturbative photon wave
function. 

This is the problem for which we propose a solution in the present
paper. A central role in our parametrization of the photon wave function
will be played by the effective quark mass. With increasing resolution
of the photon the light quarks experience a sort of chiral
transition~\cite{pol76} with constituent masses at low $Q^2$ becoming
current quark masses for $Q^2\ge 1$~GeV$^2$. The outline of the paper is
as follows. In  Sec.~\ref{sec:ho} we demonstrate that the two
dimensional harmonic oscillator can be used to model our light cone
parametrization of the photon wave function. In Sec.~\ref{sec:ph} the
approximate form of the photon wave function is given at low $Q^2$.
Section~\ref{sec:cs} deals with the calculation of the inclusive virtual
photon cross section for large energy and small $x_B$. In
Sec.~\ref{sec:fe} we discuss corrections from finite energy.

\section{The two dimensional harmonic oscillator as a model for the
photon wave function}\label{sec:ho}

In light cone perturbation theory, the photon wave function is given by 
the light cone energy denominator and spin matrix elements. Leaving
aside the spin complexities, we have the approximate form:
\begin{equation}\label{photon}
\psi_{\gamma}(z,{\bf r})\propto\int\frac{d^2{\bf k}}{(2\pi)^2}
\frac{e^{i{\bf kr}}}{{\bf k}^2+z(1-z)Q^2+m_f^2}=
\frac{1}{2\pi} K_0(\sqrt{z(1-z)Q^2+m_f^2}\;|{\bf r}|)
\end{equation}
Here $m_f$ is the current quark mass of the quark and antiquark with 
flavor $f$. The transverse extension of the wave function is given by
$r_{\perp}\sim\varepsilon^{-1}$, where
$\varepsilon=\sqrt{z(1-z)Q^2+m_f^2}$. For small values of $\varepsilon$,
however, the confining gluonic forces and/or the spontaneous chiral
symmetry breaking will intervene and limit the transverse extent of the
photon wave function. At large energy, far away from the target, the
perturbative wave function is certainly valid, but at finite small
$Q^2$, there is enough time for the photon to dress up like a bound
state. 

In quantum mechanics, the two-dimensional harmonic oscillator is
a very reasonable testing ground for the behavior of the photon wave
function in transversal space, since the harmonic oscillator has two
essential features in common with the behavior of our $q\bar{q}$ dipole
in QCD: on the one hand, large transverse distances are prohibited,
because of the harmonic potential (confinement), and, on the other hand,
the potential vanishes at the origin which corresponds in our problem to
the fact that, for short times and small relative transverse distances
of the quark and antiquark, the dynamics is entirely governed by the
kinetic energy in the hamiltonian (asymptotic freedom). The Green
function of the two dimensional harmonic oscillator
\begin{equation}\label{full}
G({\bf r},0,M)=\sum _{{\bf n}=(n_1,n_2)}
\frac{\psi_{\bf n}({\bf r})^*\psi_{\bf n}(0)}{(n_1+n_2 +1)\,\omega+M},
\end{equation}
shows the analogy to the photon wave function.
The wave function $\psi_{\bf n}(0)$ stands for the hard process of
$q \bar q$ production and $\psi_{\bf n}({\bf r})$ gives the transversal
extension. The short time restriction can be included by looking at the
dynamics for large negative values of $E=-M$, where large $M$
corresponds to the deep Euclidean region of QCD. The harmonic oscillator
Green function approaches for large negative values of $E=-M$ the free
two-dimensional Green function in quantum mechanics
\begin{equation}\label{free}
G_{\rm free}({\bf r},0,M)=\int\frac{d^2{\bf k}}{(2\pi)^2}
\frac{e^{i{\bf kr}}}{k^2/(2m)+M}=\frac{m}{\pi}K_0(\sqrt{2mM}\;|{\bf r}|).
\end{equation}
When we put $2mM=-2mE=z(1-z)Q^2+m_f^2$, we see directly the
similarity to the perturbative photon wave function. Since many exact
results are available for the harmonic oscillator, we will be able to
check several manipulations on the transverse wavefunctions. In the end,
we will not apply the results of the harmonic oscillator directly to
QCD, but we extract the essential features from the non-relativistic
model and transpose them into relativistic quantum field theory.

To begin with, let us illustrate the point made in the introduction on
the cancellation mechanism.
In diffractive vector meson production, we need the matrix element of
$r^2$ of the $q\bar q$ state with the vector meson state. In the
harmonic oscillator case, this second moment,
\begin{equation}\label{moment-def}
\left\langle r^2\right\rangle_{00}=\int d^2{\bf r}\,\,r^2\, 
G({\bf r},0,M)\,\psi_{00}({\bf r}),
\end{equation}
can be computed exactly for the full Green function. Using the spectral 
decomposition of the full Green function, Eq.~(\ref{full}), and the
one-dimensionnal harmonic oscillator property
$$
\langle n|x^2|0\rangle={\sqrt{2}\delta_{n2}+\delta_{n0}\over2\omega m},
$$
one gets
$$
\left\langle r^2\right\rangle_{00}={1\over\omega m}\left[
{\sqrt{2}\psi_{02}(0)\over3\omega+M}+{\psi_{00}(0)\over\omega+M}\right].
$$
In the last equation, the contribution of the second excited state and
the ground state are shown separately. With 
$\psi_{00}(0)=\sqrt{\omega m/\pi}$ and 
$\psi_{02}(0)=-\sqrt{\omega m/2\pi}$, they add up to
$$
\left\langle r^2\right\rangle_{00}=
\sqrt{\omega\over\pi m}{2\over(3\omega+M)(\omega+M)}.
$$
The free Green function gives for the same moment
$$
\left\langle r^2\right\rangle_{00\,\,{\rm free}}=\int d^2{\bf r}\,
r^2{m\over\pi}K_0\left(\sqrt{2mM}r\right)\psi_{00}({\bf r})
\sim{2\psi_{00}(0)\over mM^2}=\sqrt{\omega\over\pi m}{2\over M^2},
$$
As expected, the large $M$ behavior of the full Green function
moment is exactly reproduced by the free Green function. The important
lesson of this simple exercise is the demonstration that the large $M$
behavior follows from a delicate cancellation between ground and excited
state contributions. Note that the contribution of one single state is
$\sim 1/M$ and would overshoot the full result at large values of $M$.
This is to be compared with the discussion we had in the introduction.

Although the harmonic oscillator Green function, $G({\bf r},0,t)$, is
known analytically, we know of no such representation for the Fourier
transformed $G({\bf r},0,M=-E)$; therefore, we have obtained an
``exact'' expression by performing the sum in Eq.~(\ref{full}) with the
first 500 terms using the known wave functions of the harmonic
oscillator. A simple calculation shows, that even for moderate values of
$M$, say $5\omega$, one needs more than twenty intermediate terms in the
representation Eq.~(\ref{full}) in order to obtain a better accuracy
than the one from the free Green function. In terms of vector dominance,
this means that we need many intermediate vector mesons in order to get
an adequate description for moderate values of the photon virtuality. As
we will show, a much more efficient procedure is to shift the argument
$2mM$ in the free Green function by a $M$ dependent value $s_0$; it
turns out that this method gives, even for $M=0$, a very decent
approximation to the full Green function. 

Since the exact Green functions is available for the harmonic
oscillator, the shift parameter can be calculated by comparing the
modified free Green function with the exact one. This procedure cannot
be, however, carried over to the lightcone wave function of the photon.
We consider, therefore, the ``two-point'' function, $\Pi(M)$, and its
derivatives 
\begin{equation}
\Pi^{(n)}(M):=(-1)^n \frac{d^n}{dM^n}\Pi(M):=
(-1)^n\frac{d^n}{dM^n}G(0,0,M),
\end{equation}
instead of the ``three-point'' functions $G({\bf r},0,M)$. For
convergence, we need at least one derivative. The two-point functions
have been extensively studied in QCD, especially with sum rule
techniques~\cite{shi79}. From this, we know that in an asymptotically
free theory the ansatz ``one resonance plus perturbative continuum'' is
a very good phenomenological representation for the two-point function
in the Euclidean region. Our method is then to make for the two-point
function, $\Pi^{(n)}(M)$, the model ``one resonance plus perturbative
continuum'' and to use for the three-point function an approximate form
which can be parametrized easily and adjusted in such a way that the
two-point function obtained from it agrees with the model two-point
function. 

For the derivatives of $\Pi(M)$, the ansatz ``one resonance plus 
perturbative continuum'' reads, to lowest order,
\begin{equation}\label{model}
\Pi^{(n)}_{\rm ph}(M)=\frac{n! |\psi_{00}(0)|^2}{(\omega+M)^{n+1}}
+\frac{m}{2 \pi} \frac{(n-1)!}{(s_t/2m+M)^n},
\end{equation}
where $\psi_{00}$ is the ground state wave function and $s_t$ the
continuum threshold above which we use the perturbative Green function. 
Duality states that the integral from 0 to $s_t$ over the imaginary part 
of the free (i.e., perturbative) two-point function accounts for $\pi$
times the residue at the resonance pole:
$$
\int_0^{s_t}dk^2\,{\rm Im}\,\Pi_{\rm free}(k^2)=\pi|\psi_{00}(0)|^2.
$$
Thereby, we obtain the continuum threshold
$$
s_t=4\pi|\psi_{00}(0)|^2=4\omega m.
$$
 
As a simple approximate Green function, we consider the free Green
function with a shifted argument
\begin{equation}\label{approx1}
G_{\rm a}({\bf r},0,M,s_0)=\frac{m}{\pi} K_0(\sqrt{2mM+s_0}\,r).
\end{equation}
The derivatives of the correponding approximate two-point function have 
the form
\begin{equation}\label{approx2}
\Pi^{(n)}_{\rm a}(M,s_0)=(-1)^n\frac{\partial^n}{\partial M^n} 
G_{\rm a}(0,0,M,s_0)=\frac{m}{2\pi}\frac{(n-1)!}{(s_0/2m+M)^n}.
\end{equation}
(Notice that in the left hand side the relation involves the partial 
derivative and not the total derivative.) By equating the approximation 
Eq.~(\ref{approx2}) to the phenomenological function Eq.~(\ref{model})
\begin{equation}\label{approx3}
\Pi^{(n)}_{\rm a}(M,s_0)=\Pi^{(n)}_{\rm ph}(M),
\end{equation}
we can determine the only free parameter of the approximation, namely 
$s_0(M)$, and obtain
\begin{equation}\label{s0-def}
s_0/2m=(\omega+M)^{(n+1)/n} (2\omega+M)
[2n\omega(2\omega+M)^n+(\omega+M)^{n+1}]^{-1/n}-M.
\end{equation}
The exact form of the shift depends on the number of differentiations 
assumed. Yet, at $M=0$, $s_0/2m$ is around $\omega/2$ and decreases to 
become a small correction to $M$, i.e., less than 5\%, for $M$ larger 
than $3\omega$.

\begin{figure}[ht]
$$\vbox to 95mm{\vfill\smash{\epsfysize=105mm\epsfbox{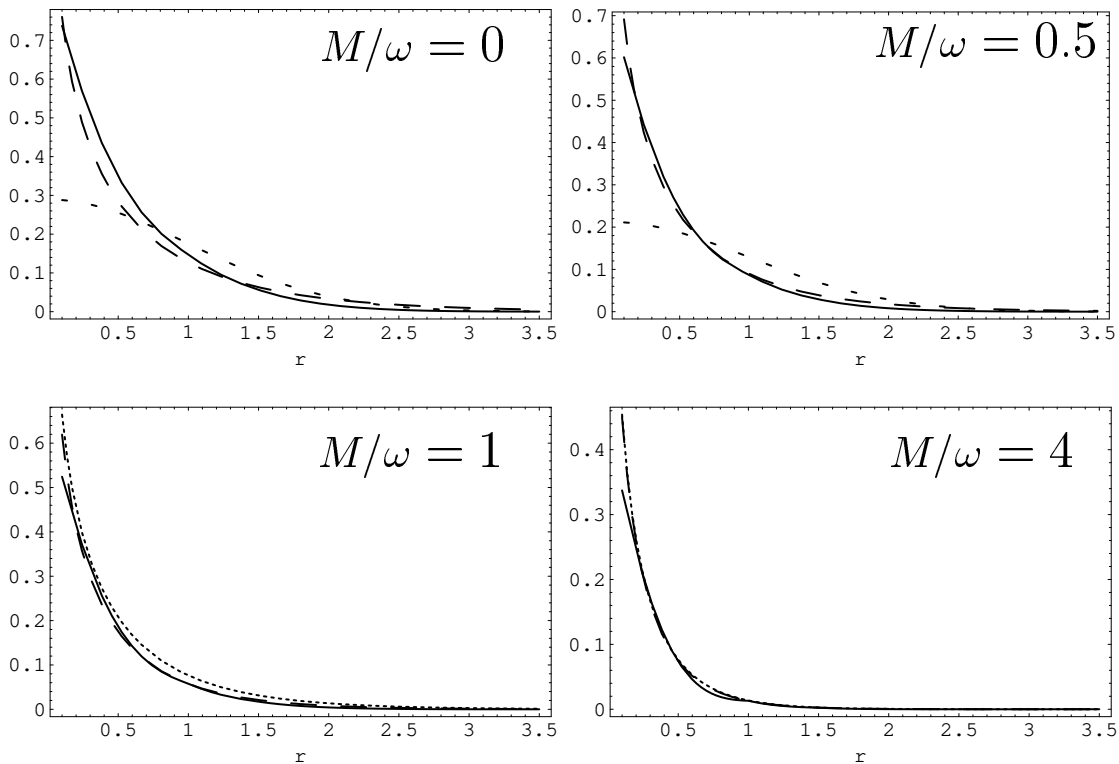}
}\vskip-8mm}$$
\caption{Green functions (in units of $m$) of an harmonic oscillator as 
function of $r$ (in units of $1/\protect\sqrt{\omega m}$) for different 
values of $M/\omega$. Solid line: exact Green function $G({\bf r},0,M)$,
long dashes: our approximation $G_{\rm a}({\bf r},0,M,s_0)$, i.e., the
shifted free Green function Eq.~(\protect\ref{approx1}) with the shift
$s_0$ from Eq.~(\protect\ref{s0-def}); short dashes: approximation with
two resonances, dots: free Green function.}
\label{h-o} 
\end{figure}

In Fig.~\ref{h-o}, we display the approximated Green function obtained 
with a $n=3$ shift (long dashes) and the full Green function (full
line), for $M/\omega=0$, 0.5, 1 and 4. For the first two values, we also
show for comparison a two-resonance approximation to Eq.~(\ref{full}).
For the last two values, the short dashes represent the free Green
function;  for $M/\omega=4$ the approximated Green function can hardly
be distinguished from the free one.   

We estimate the quality of our approximation by forming the moments
$$
\left\langle r^n\right\rangle:=\int d^2{\bf r}\,\,r^n\,G^2({\bf r},0,M),
$$
which can be evaluated easily, both for the full and the approximate 
Green function. Since in electroproduction, for high virtualities, the 
$q\bar q$-proton cross section behaves approximately like $r^2$ and, in
the hadronic region, like $r^{1.5}$~\cite{dos97}, we pay special
attention  
to the moment $\langle r^2\rangle$. In Table~\ref{moment}, we give the 
relative differences of the exact and approximated moments
\begin{eqnarray*}
\Delta_2&:=&(\langle r^2\rangle-\langle r^2\rangle_{\rm a})/
\langle r^2\rangle,\\
\Delta_0&:=&(\langle r^0\rangle-\langle r^0\rangle_{\rm a})/
\langle r^0\rangle,
\end{eqnarray*}
for different values of $M$ and numbers of differentiations $n$. As can
be expected, the lower $n$-value, $n=1$, yields a good approximation for
the zeroth moment, whereas the second moment is well reproduced with
$n=3$. The maximal error for $\langle r^2\rangle$ turns out to be 11\%
(at $M$=0).

\begin{table}[ht]
\begin{center}
\begin{tabular}{|l|c|c|c|c|c|c|}
\hline
$M/\omega$&\multicolumn{2}{c|}{$n=1$}&\multicolumn{2}{c|}{$n=3$}&
\multicolumn{2}{c|}{$n=5$}\\
\hline
& $\Delta_2$ & $\Delta_0$& $\Delta_2$ & $\Delta_0$& $\Delta_2$ & $\Delta_0$\\
\hline 
0 & 1.09 & 0.01 &0.11 &-0.26 & -0.16 & -0.35 \\
0.5&0.51&0.02&0.05&-0.14&-0.13&-0.23\\
1&0.28&0.02&0.02&-0.09&-0.11&-0.15\\ \hline
\end{tabular}
\end{center}
\caption{Differences of the exact and approximated moments:
$\Delta_2:=(\langle r^2\rangle-\langle r^2\rangle_{\rm a})/\langle r^2\rangle$ 
and
$\Delta_0:=(\langle r^0\rangle-\langle r^0\rangle_{\rm a})/\langle r^0\rangle$}
\label{moment}
\end{table}

\begin{figure}[ht]
$$\epsfxsize=11cm\epsfbox{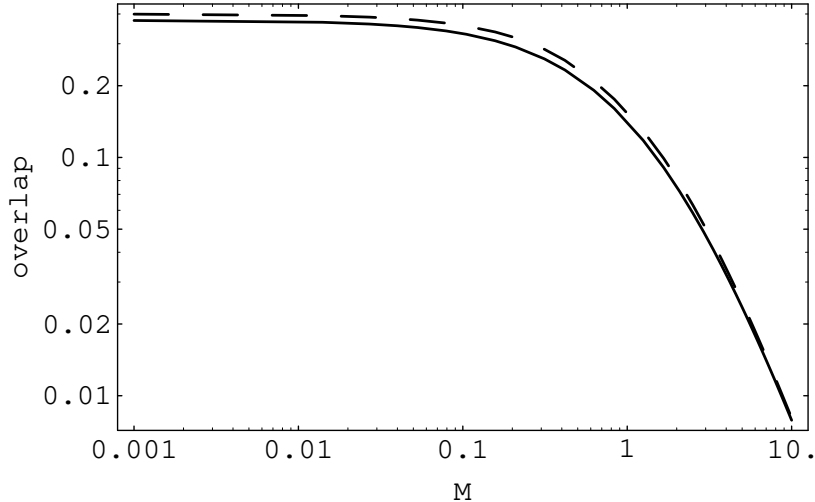}$$
\caption{Overlap Eq.~(\protect\ref{moment-def}) of the Green function of 
a two-dimensional harmonic oscillator with the ground state; solid line: 
exact result, dashes: calculated with our approximated Green function
from Eq.~(\protect\ref{approx1}) (in units $m=\omega=1$).}
\label{h-o-moment} 
\end{figure}

We also compare the overlap of the exact Green function and the ground 
state wave function, i.e., $\left\langle r^2\right\rangle_{00}$ defined 
in Eq.~(\ref{moment-def}), with the respective overlap including the 
approximated Green function
$$
\left\langle r^2\right\rangle_{00\,{\rm a}}=\int d^2{\bf r}\,\,r^2\, 
G_{\rm a}({\bf r},0,M)\psi_{00}({\bf r}).
$$
This is shown in Fig.~\ref{h-o-moment} and Table~\ref{00-moment}. One
sees that the shifted free Green function gives an good estimate of the
exact matrix element, the relative error being at most 10\%.

\begin{table}[ht]
\begin{center}
\begin{tabular}{|c|l|l|l|}\hline
$M$& approx& exact&error\\ \hline
0&           0.400  &  0.376  &   0.06\\
1&           0.155  &  0.141  &   0.10\\
2&           0.0819 &  0.0752 &   0.09\\
3&           0.0504 &  0.0470 &   0.07\\
4&           0.0342 &  0.0322 &   0.06\\
5&           0.0247 &  0.0235 &   0.05\\
6&           0.0186 &  0.0179 &   0.04\\
7&           0.0146 &  0.0141 &   0.03\\
8&           0.0117 &  0.0114 &   0.03\\
9&           0.0096 &  0.0094 &   0.03\\
10&          0.0081 &  0.0079 &   0.02\\ \hline
\end{tabular}
\end{center}
\caption{Comparison of exact and approximated overlap with ground state
(in units $m=\omega=1$).}\label{00-moment}
\end{table}

In Table~\ref{shifts}, we compare the shifts $s_0$ obtained from
Eq.~(\ref{s0-def}) with $n=$2, 3 and 4 differentiations, with the ones
needed to get exact agreement for the second moment. To reproduce the
second moment, the displacement $s_0$ is quite different from the value
of $s_0$ which one obtains from Eq.~(\ref{approx3}). Our method
underestimates the shifts considerably for large $M$, but the overall
errors on the matrix elements remain small as shown in
Table~\ref{00-moment} and Fig.~\ref{h-o-moment}. Indeed we already
noticed that for $M\gg\omega$ the full Green function Eq.~(\ref{full})
is well approximated by the free one Eq.~(\ref{free}), i.e., $s_0$ can
be safely set equal to 0 in this region.

\begin{table}[ht]
\begin{center}
\begin{tabular}{|c|c|c|c|c|c|c} \hline
$M/\omega$&  $n=2$&    $n=3$&    $n=4$&    exact\\ \hline
0&           0.485&    0.547&    0.593&    0.585\\
1&           0.279&    0.344&    0.397&    0.455\\
2&           0.179&    0.234&    0.283&    0.385\\
3&           0.123&    0.168&    0.211&    0.325\\
4&           0.090&    0.127&    0.162&    0.27\\
5&           0.068&    0.099&    0.129&    0.255\\ \hline
\end{tabular}
\end{center}
\caption{$s_0/2m$ determined by the sum rule method for $n=$2, 3, 4 
subtractions and adjusted to give exact overlap.}\label{shifts}
\end{table}

\section{Approximate photon wave function extended to low values of $Q^2$}
\label{sec:ph}

We now want to apply the approximation methods developped for the
harmonic oscillator to the photon wave function. For this purpose, 
we consider first the polarization tensor for the vector current 
$J_\mu=\bar{\psi}\gamma^\mu \psi$ of a quark of mass $m$
\begin{eqnarray*}
\Pi^{\mu\nu}(q,m^2)&=&
\int d^4x\,e^{iqx}\langle 0|T\left[J^\mu(x)J^\nu(0)\right]|0\rangle\\
&=&(q^\mu q^\nu-g^{\mu\nu} q^2)\,\Pi(q^2,m^2).
\end{eqnarray*}
At large $q^2$, the imaginary part of the polarization function 
$\Pi(q^2,m^2)$ is obtained to lowest order in perturbation theory from 
the free quark-antiquark propagation. One has
\begin{equation}\label{imaginary}
{\rm Im}\,\Pi(q^2,m^2)=\frac{N_c}{12\pi}\frac{q^2+2 m^2}{q^2} 
\sqrt{1-\frac{4 m^2}{q^2}}.
\end{equation}
The polarization function itself is only determined up to a subtraction
constant, but its derivatives
\begin{equation}
\Pi^{(n)}(Q^2=-q^2,m^2):=\frac{\partial^n}{\partial(q^2)^n}\Pi(q^2,m^2),
\end{equation}
can be written for $n\ge 1$ by dispersion relations:
\begin{equation}\label{dispersion}
\Pi^{(n)}(Q^2,m^2)=\frac{n!}{\pi}\int_{4 m^2}^{\infty}ds
\frac{{\rm Im}\,\Pi(s,m^2)}{(s+Q^2)^{n+1}}.
\end{equation}

Due to asymptotic freedom, the polarization tensor has a very good  
representation in the Euclidean region consisting of the ground state 
vector meson with mass $m_V$ and residue $F_V$ and perturbative $q\bar q$ 
continuum calculated with current quark masses:
$$
\Pi^{(n)}_{\rm ph}(Q^2)={n!\,F_V^2\over (Q^2+m_V^2)^{n+1}}
+\frac{n!}{\pi} \int_{s_t}^{\infty}ds\frac{{\rm Im}\,\Pi(s,m_f^2)}
{(s+Q^2)^{n+1}}.
$$
$V$ is the lowest lying vector meson in the flavor channel considered, i.e., 
$\rho,\omega=(u\bar{u}\mp d\bar{d})/\sqrt{2}$ and $\phi=s\bar{s}$. 
$F_V$ is the decay constant of this vector meson defined through 
$$
F_V m_V\varepsilon^\mu(q,\lambda)=\langle 0|J^\mu(0)|V(q,\lambda)\rangle.
$$
The continuum threshold $s_t$ can be related to the decay constant $F_V$ 
by local duality:
$$
F_V^2=\frac{1}{\pi}\int_{4m_f^2}^{s_t}ds\,{\rm Im}\,\Pi(s,m_f^2).
$$

We copy now the procedure from the discussion of the harmonic oscillator.
The photon wave function plays the role of the three point function which 
we want to approximate. As in Eq.~(\ref{approx1}), we take as approximate 
Green function the free Green function but shift the variable $Q^2$.
The structure of the perturbative photon wave function is of the form 
(see Eq.~(\ref{photon})):
$$
\psi_\gamma\propto K_0\left(\sqrt{z(1-z)Q^2+m_f^2}\, r\right).
$$
A $Q^2$ dependent shift thus corresponds to a replacement of the current 
mass $m_f$ by an effective mass $m_{\rm eff}(Q^2)$. Here some improvement 
is still possible, since in reality the virtuality $Q^2$ appears in 
combination with the light cone momentum fraction through the term $z(1-z)$. 
We leave this difficulty aside for the moment and shall return to it 
later. From Eqs.~(\ref{imaginary})--(\ref{dispersion}), 
we form the derivatives of the approximate polarization function
\begin{eqnarray}\label{polarization}
&&\Pi^{(n)}_{\rm a}(Q^2,m_{\rm eff}^2):=
{\partial^n\over\partial(-Q^2)^n}\Pi_{\rm a}(Q^2,m_{\rm eff}^2)\\
&&={N_c\over 12\pi^2}{\partial^n \over\partial(-Q^2)^n}\left[
-{4 m_{\rm eff}^2\over Q^2}-\left(1-{2m_{\rm eff}^2\over Q^2}\right)
\sqrt{1+{4m_{\rm eff}^2\over Q^2}}\ln{\sqrt{1+4m_{\rm eff}^2/Q^2}
+1\over\sqrt{1+4m_{\rm eff}^2/Q^2}-1}\right]\nonumber.
\end{eqnarray}
Next, in complete analogy to Eq.~(\ref{approx3}), we determine the 
effective mass in such a way that the shifted two point function, 
$\Pi^{(n)}_{\rm a}$, is equal to the model two point function 
$\Pi^{(n)}_{\rm ph}$:
\begin{equation}\label{m_eff-def}
\Pi^{(n)}_{\rm a}(Q^2,m_{\rm eff}^2)=
\frac{n!\,F_V^2}{(Q^2+m_V^2)^{n+1}}+\frac{n!}{\pi} 
\int_{s_t}^{\infty}ds\frac{{\rm Im}\Pi(s,m_f^2)}{(s+Q^2)^{n+1}}.
\end{equation}

\begin{figure}[ht]
$$\epsfxsize=10cm\epsfbox{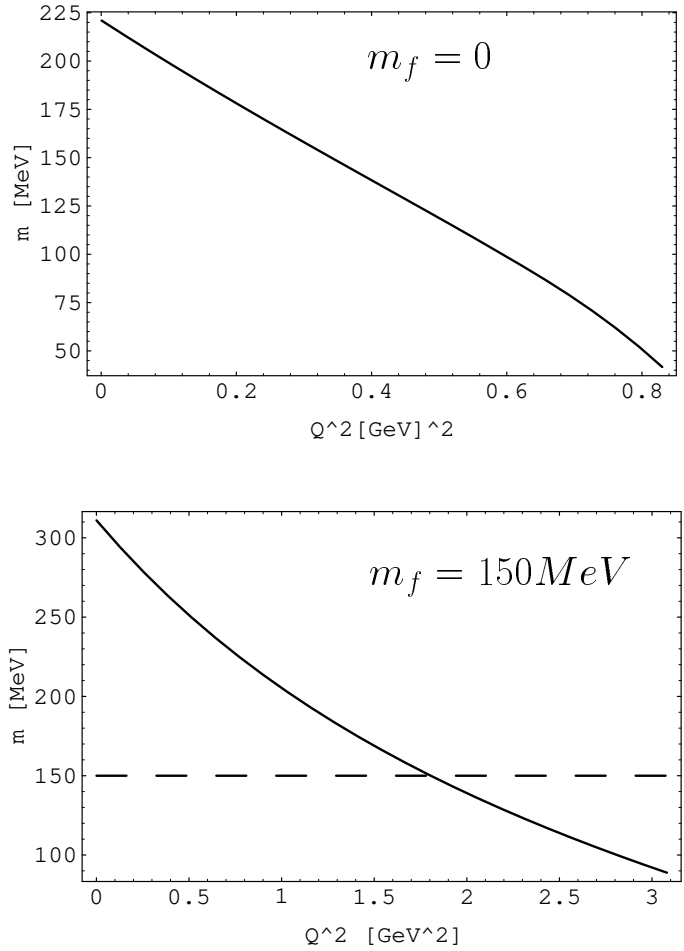}$$
\caption{Effective quark masses which reproduce the model polarization
function, see Eq.~(\protect\ref{m_eff-def}). (a) The $Q^2$ behavior for
light quarks and (b) strange quarks.}\label{quark-mass} 
\end{figure}

\begin{figure}[ht]
$$\epsfxsize=10cm\epsfbox{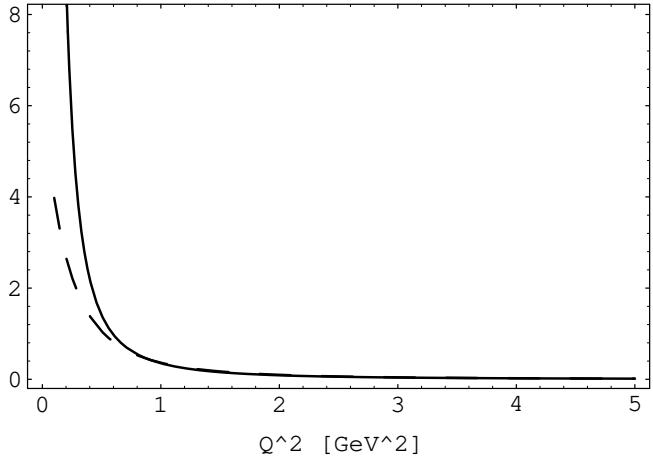}$$
\caption{Second derivative of the polarization function; dashes: model 
function $\Pi_{\rm ph}(Q^2)$, i.e., one resonance plus continuum; solid:
lowest order perturbative expression with zero mass. Note for
$Q^2>1$~GeV$^2$ the perturbative expression becomes indistinguishable
from the perturbative one.}\label{pi2a} 
\end{figure}

The method gives the effective quark mass as a function of $Q^2$ on the
left hand side from the purely hadronic parameters $F_V$ and $m_V$ on
the right hand side. In Fig.~\ref{quark-mass}, we display the resulting
effective quark masses $m_{\rm eff}(Q^2)$, putting $n=2$, for massless
current quarks and strange quarks with a current mass $m_s=150$~MeV. As
for the harmonic oscillator, the result depends on the number of
differentiation assumed. For $m_q=0$, the effective mass starts at
$Q^2=0$ at $220$~MeV for $n=2$ and $245$~MeV for $n=3$ which are typical
constituent quark mass values. The effective mass drops to 0 at 
$Q_0^2\approx 1.5$--$2\,m_\rho^2$. Beyond this value of $Q^2$, it
formally becomes imaginary. We have seen in the harmonic oscillator that
the method underestimates the shift at large virtuality, but the errors
introduced are less than 5\%. We therefore put the effective mass equal
to zero above $Q_0^2$. To be specific, we use the simple linear
parametrization 
\begin{eqnarray}
m_{\rm eff}(Q^2)&=&.22\,(1.-Q^2/Q_0^2),\hbox{ in GeV, for 
$Q^2\le Q_0^2=1.05$~GeV$^2$,}\nonumber\\
m_{\rm eff}(Q^2)&=&0,\hbox{ for $Q^2\ge Q_0^2$.}\label{effective-mass}
\end{eqnarray}
To support our point to set the effective light quark mass to zero for 
large $Q^2$, we display in Fig.~\ref{pi2a} the second derivative of the 
model Green function together with the free one. We see that the two 
agree at large $Q^2$ values where our effective mass formally would 
become imaginary. A more refined procedure would be to make a smooth 
connection between the small $Q^2$ behavior obtained with the present 
method and the behavior obtained around 1~GeV$^2$ using Operator 
Product Expansion~\cite{pol76}. We shall not dwell on this possibility 
in the following. For the strange quark, we refer to the second part of 
Fig.~\ref{quark-mass}, which shows that the corresponding starting value 
for the constituent strange mass is $310$~MeV. It reaches its asymptotic 
value in the range $Q^2=1.5$--2~GeV$^2$. A simple parametrization for 
the strange quark mass is 
\begin{eqnarray}
m_{s{\rm eff}}(Q^2)&=&.15+.16\,(1.-Q^2/Q_0^2),\hbox{ in GeV, for 
$Q^2\le Q_0^2=1.6$~GeV$^2$,}\nonumber\\
m_{s{\rm eff}}(Q^2)&=&.15,\hbox{ for $Q^2\ge Q_0^2$.}
\label{effective-s-mass}
\end{eqnarray}

\begin{figure}[ht]
$$\epsfxsize=10cm\epsfbox{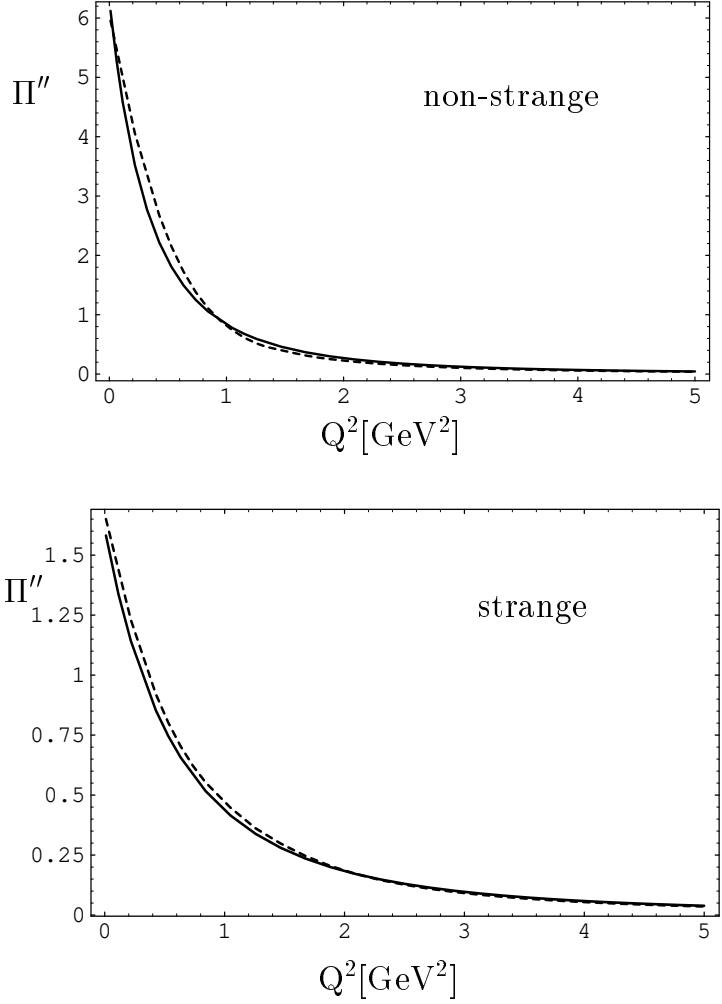}$$
\caption{Second derivative of the polarization function for vector
current of light and strange quarks. Solid: model function 
$\Pi_{\rm ph}(Q^2)$, dashes: approximate expression 
$\Pi_{\rm a}(Q^2,m_{\rm eff})$ with the quark mass depending on
$Q^2_{\rm eff} =4z(1-z)Q^2$. Note that for the quark mass depending on
$Q^2$ alone we get exact agreement by construction.}\label{pi2b} 
\end{figure}

As mentioned before, in the photon wave function, $Q^2$ appears together 
with the factor $z(1-z)$. We re-analyzed the polarization function
$\Pi^{(n)}_{\rm a}$ including an effective quark mass as a function of 
$Q^2_{\rm eff}=4z(1-z)\,Q^2$. This modifies the treatment that leads to 
Eq.~(\ref{polarization}) along the following lines. The imaginary part 
of the polarization function $\Pi(q^2,m^2)$ of the vector two point 
function $\Pi_{\mu\nu}$ can be written as
$$
{\rm Im}\,\Pi(q^2,m^2)=\frac{N_c}{12\pi}\int_0^1dz\,\frac{q^2+2m^2}{q^2}\,
\Theta\Big[z(1-z)q^2-m^2\Big].
$$
Upon integration over $z$, it yields to the familiar result given in
Eq.~(\ref{imaginary}). We then obtain the derivatives by dispersion 
relations:
\begin{equation}\label{dispersion2}
\Pi^{(n)}(Q^2,m^2)=\frac{N_c}{12\pi^2}\int_0^1dz\,
\int_{m^2/(z(1-z))}^\infty ds\,\frac{s+2m^2}{s}\frac{n!}{(s+Q^2)^{n+1}}.
\end{equation}
Setting $m=m_{\rm eff}(Q^2_{\rm eff})$ in the right hand side of this 
equation gives a new expression for $\Pi^{(n)}_{\rm a}$. We numerically 
find that the phenomenological two point function can be reproduced with 
an accuracy better than 10\% (see Fig.~\ref{pi2b}) with parametrizations 
similar to that given above, provided we change the value of $Q_0^2$. We 
get for light quarks
\begin{eqnarray}
m_{\rm eff}(Q^2_{\rm eff})&=&.22\,(1.-Q^2_{\rm eff}/Q'^2_0),
\hbox{ in GeV, for $Q^2_{\rm eff}\le Q'^2_0=0.69$~GeV$^2$,}\nonumber\\
m_{\rm eff}(Q^2_{\rm eff})&=&0,\hbox{ for $Q^2_{\rm eff}\ge Q'^2_0$.}
\label{z-dependent-mass}
\end{eqnarray}
The lower scale $Q'^2_0$ is due to the difference between the average 
in $z$ occuring in Eq.~(\ref{dispersion2}) and the value at the average 
$z=1/2$. For the strange quark, the result is
\begin{eqnarray}
m_{s{\rm eff}}(Q^2_{\rm eff})&=&.15+.16\,(1.-Q^2_{\rm eff}/Q'^2_0),
\hbox{ in GeV, for $Q^2_{\rm eff}\le Q'^2_0=1.16$~GeV$^2$,}\nonumber\\
m_{s{\rm eff}}(Q^2_{\rm eff})&=&.15,\hbox{ for $Q_{\rm eff}^2\ge Q'^2_0$.}
\label{z-dependent-s-mass}
\end{eqnarray}

\section{Inclusive photon-proton cross section}\label{sec:cs}

Let us now consider the forward Compton amplitude, i.e., the amplitude
in Eq.~(\ref{amplitude}), with the replacement
$\psi^*_V\to\psi^*_\gamma$, at $t=0$:
$$
{\cal M}(\gamma^* p\to\gamma^* p)=is\int {dzrdr\over 2}
|\psi_\gamma(z,r)|^2\,J_p(z,r),
$$
where we use the notation $J_p(z,r)=J_p^{(0)}(z,r,t=0)$ for short. We
employ the same dipole-proton cross section as in our previous work on
exclusive vector meson production~\cite{dos97}, which is based on the
evaluation of gluon field strength correlators between Wilson areas
mapped out by the color neutral dipole and the proton. The absolute size
and $z,r$ dependences of the cross section $J_p(z,r)$ are determined by
the gluon condensate $\langle g^2 FF\rangle = 2.49~$GeV$^4$, the
correlation length $a=0.346$~fm and the transverse radius of the proton
$R_{\bot p}=0.52$~fm, together with the form of the correlators assumed
there.

We compute the square of the photon wave function using the expression 
given in Ref.~\cite{dos97}. This leads to the longitudinal and
transverse amplitudes
\begin{equation}\label{L-T-amplitude}
{\cal M}_{L/T}=is\sum_f\sigma_{f\,L/T}=is\sum_f e_f^2
\int_0^1 dz\int_0^{\infty}rdr {\cal I}_{L/T}(z,r)
\end{equation}
with the integrands
\begin{eqnarray}
{\cal I}_L(z,r)&=&{N_c\over4\pi^2}4z^2(1-z)^2Q^2
K_0(\varepsilon r)^2\,J_p(z,r),\\\label{T-integrand}
{\cal I}_T(z,r)&=&{N_c\over4\pi^2}\left\{[z^2+(1-z)^2]
\varepsilon^2K_1(\varepsilon r)^2+m_f^2K_0(\varepsilon r)^2\right\}
J_p(z,r).
\end{eqnarray}
The extension parameter of the photon is
$\varepsilon^2=z(1-z)Q^2+m_f^2$. It depends on the quark flavor through
the quark mass and thus each flavor contributes in a different way to
the above sums. $K_0$, $K_1$ are modified Bessel functions. They arise
from the perturbative light cone wave function of the photon when one
takes into account all the spin complexities ignored in
Sec.~\ref{sec:ho}. For a general dipole-proton cross section, $J_p$,
these expressions are identical to those given in
Refs.~\cite{bjo71,nik91}. Our dipole-proton cross section $J_p(z,r)$ 
calculated from gluon-gluon correlators describes the scattering on a
proton of a loop with transverse size $r$ and infinite extension along
the light cone. It depends only very weakly on the longitudinal
momentum fraction $z$ and shows for values of $r$ smaller than
approximately $2a\approx 0.7$~fm the dipole behavior 
\begin{equation}\label{dipole-behavior}
J_p(z,r) \approx C r^2 \mbox{ with } C=4.3
\end{equation}
For very large values of $r$ it increases linearly and around 1~fm it
goes approximately like $r^{1.5}$ .

One can get a rough idea on the $Q^2$ dependence of the cross sections 
if one assumes the dipole behavior given in Eq.~(\ref{dipole-behavior}).
Then one obtains analytical expressions for the cross sections for each
flavor channel, setting $u=4m^2/Q^2$: 
\begin{eqnarray*}
\sigma_{f\,L}&=&{\alpha_{em}\hat{e}_f^2}{N_c\,C\over3\pi Q^2}\left[
4-{u(4+3u)\over(1+u)^{3/2}}\ln{(1+\sqrt{1+u})^2\over u}+{2u\over 1+u}
\right],\\
\sigma_{f\,T}&=&{\alpha_{em}\hat{e}_f^2}{N_c\,C\over3\pi Q^2}\left[
-4+{2\over 1+u}+{1\over\sqrt{1+u}}\left(4+2u+{u\over 1+u}\right)
\ln{(1+\sqrt{1+u})^2\over u}\right].
\end{eqnarray*}

{}From these expressions, we deduce easily the behavior for large and
small $Q^2$:

1) large $Q^2$:
\begin{eqnarray} \label{largeQ}
\sigma_{f\,L}&=&{\alpha_{em}\hat{e}_f^2}{4N_c\, C\over3\pi Q^2}
\left[1-\frac{4 m^2}{Q^2}\ln\frac{Q^2}{m^2}+
O\left(\frac{m^2}{Q^2}\right)\right],\\
\sigma_{f\,T}&=&{\alpha_{em}\hat{e}_f^2}{4N_c\, C\over3\pi Q^2}
\left[\ln\frac{Q^2}{m^2}-\frac{1}{2}
+O\left(\frac{m^2}{Q^2}\ln\frac{Q^2}{m^2}\right)\right];
\end{eqnarray}

2) small $Q^2$:
\begin{eqnarray}\label{smallQ}
\sigma_{f\,L}&=&{\alpha_{em}\hat{e}_f^2}{2N_c\, C\over45\pi m^2} 
\frac{Q^2}{m^2}\Big[1+O(Q^2/m^2)\Big],\\
\sigma_{f\,T}&=&{\alpha_{em}\hat{e}_f^2}{7N_c\, C\over9\pi m^2}
\Big[1+O(Q^2/m^2)\Big].
\end{eqnarray}

For large values of $Q^2$ and longitudinal photons the small dipole
sizes are dominant and the use of a purely perturbative photon wave
function is justified. We see indeed that the limit of the longitudinal
cross section for large values of $Q^2$ is unproblematic. For
transversal photons however small values of the longitudinal momentum
fractions $z$ and $(1-z)$ become more important and large dipole
sizes may play an important role even if the value of $Q^2$ is large.
A $r^2$ behavior of $J_p(z,r)$ thus leads to a logarithmic
divergence in the quark mass, as can be seen from Eq.~(\ref{largeQ}).
For large values of $r$ such a behavior is however unrealistic, and any
reduced increase of the form $r^{2-\epsilon}$ with $\epsilon>0$ leads to
a finite cross section even for a perturbative photon wave function and
a quark mass equal to zero. This feature stresses the importance of a
realistic model both for the long distance part of the photon wave
function and the dipole cross sections. A good $q\bar q$ wave function
and a realistic dipole-proton cross section become even more relevant
for low values of $Q^2$. Our model for the hadronic scattering part is
inherently nonperturbative. For the photon we absorbe the
nonperturbative effects in a virtuality dependent constituent mass. We
stress that all input to these models is taken from sources outside the
realm of electroproduction.   

\subsection{Photoproduction}

Let us compare our computed cross section with data, starting with 
photoproduction. In the following we consider data at a center of 
mass energy $W=20$~GeV. We choose this value because it is the one 
where the model parameters are adjusted to fit the corresponding 
proton-proton elastic scattering data. At this center of mass energy, 
the photon-proton total cross section is $118~\mu$b. The Pomeron part 
of the Donnachie-Landshoff fit~\cite{don94} gives 
$\sigma_{\rm Pom}=110~\mu$b, the remaining part being attributed to
other Regge trajectories. The model of the stochastic vacuum accounts
for the Pomeron part of the cross section. The Reggeon contribution to
strange (and heavier) quark interaction is much suppressed (Zweig
suppression) and can be neglected. The charm quark contribution is
measured independently and is rather small, 1~$\mu$b. In the present
study, we focus on $u$, $d$ and $s$ which give the bulk of the cross
section. According to Donnachie and Landshoff, the strange quark
contribution to the total cross section is $8.3~\mu$b. 

Our theoretical results for the light, $u$ and $d$, and strange quark 
contributions are: 
\begin{eqnarray*}
\sigma_{u+d}&=&84~\mu{\rm b},\\
\sigma_s&=&9.4~\mu{\rm b}.
\end{eqnarray*}

Some comments are called for. To illustrate our purpose, it is 
again useful to stick to the approximate amplitudes derived 
by assuming the short distance behavior Eq.~(\ref{dipole-behavior}). 
In Eq.~(\ref{dipole-behavior}) the magnitude of the cross section, 
given by $C\approx 4.3$, is determined by the properties of the QCD 
vacuum, namely the gluon condensate and the correlation length of 
gluon field strength correlators, and the proton radius. As we have 
seen in the beginning of this section, the transverse extension of 
the photon is bounded by the value of the quark mass, 
$r\le 1/m_{\rm eff}$, therefore the $r^2$ dependence of the cross 
section is certainly correct if the quark mass is large enough, 
otherwise it gives a first approximation. The limit of small $Q^2$ was
given in Eq. (\ref{smallQ}):
\begin{equation}\label{photoproduction}
\sigma_f=\alpha_{\rm em}\hat{e}_f^2{7N_c C\over9\pi m_f^2}.
\end{equation}
This means that the photoproduction cross section depends crucially 
on the value of the constituent quark mass. The reproduction of the
cross section within 15 \% is encouraging especially taking into account
the fact that we have determined the effective quark mass,
Eq.~(\ref{effective-mass}), without any recourse to  electroproduction 
phenomenology. If we attributed the remaining 15 \% difference between
the experimental value of the cross section and our theoretical one to
the value of $m_{\rm eff}$, we would get a 8 \% decrease of the
effective quark mass value. Due to other sources of uncertainty, this 
refinement does not make sense here. We also notice that, in the 
approximate cross section, 40\% come from the second term in 
Eq.~(\ref{T-integrand}) which is proportional to the quark mass squared.
Within the present determination of $m_{\rm eff}$, this supports the
interpretation of the modification of the photon extension parameter as
being due to the generation of the effective quark mass rather than
being just a shift in the argument of $K_1$ in Eq.~(\ref{T-integrand}).
For the strange quark, the comparison with the extracted cross section
is correct within 10\%. As an indication, we note that our amplitudes
with a current strange quark mass of 150~MeV would produce a much too
big cross section of $31~\mu{\rm b}$.

\subsection{Electroproduction}

We now consider virtual photon, $Q^2\ne 0$, scattering off a proton. 
We form the structure functions 
\begin{eqnarray}
F_2(Q^2)=\frac{Q^2}{\pi e^2}(\sigma_T+\sigma_L)\\ 
F_L(Q^2)=\frac{Q^2}{\pi e^2}\sigma_L.
\end{eqnarray}
Since the light and strange quarks contribute in a different way, we 
first calculate the corresponding quantities separately. Special 
attentions will be paid to the $Q^2$ dependence of the structure 
functions at fixed $W=20$ GeV, which corresponds to the energy 
where we determined our input dipole-proton cross section. 
We want to investigate the question whether the chiral transition from
the constituent quark to the parton can be seen in the inclusive
electron scattering data. In our theoretical calculation the effective
quark mass $m_{\rm eff}(Q^2)$ evolves with the photon virtuality $Q^2$.
A priori it is not clear whether the photon virtuality or the
combination of light cone momenta and $Q^2$, namely 
$Q^2_{\rm eff}=4z(1-z)Q^2$, should be used in the running of the quark 
mass for low virtualities. In the second case the integration over the 
light cone momentum fraction $z$ may wash out the chiral transition
effect.

\begin{figure}
$$\epsfxsize=13cm\epsfbox{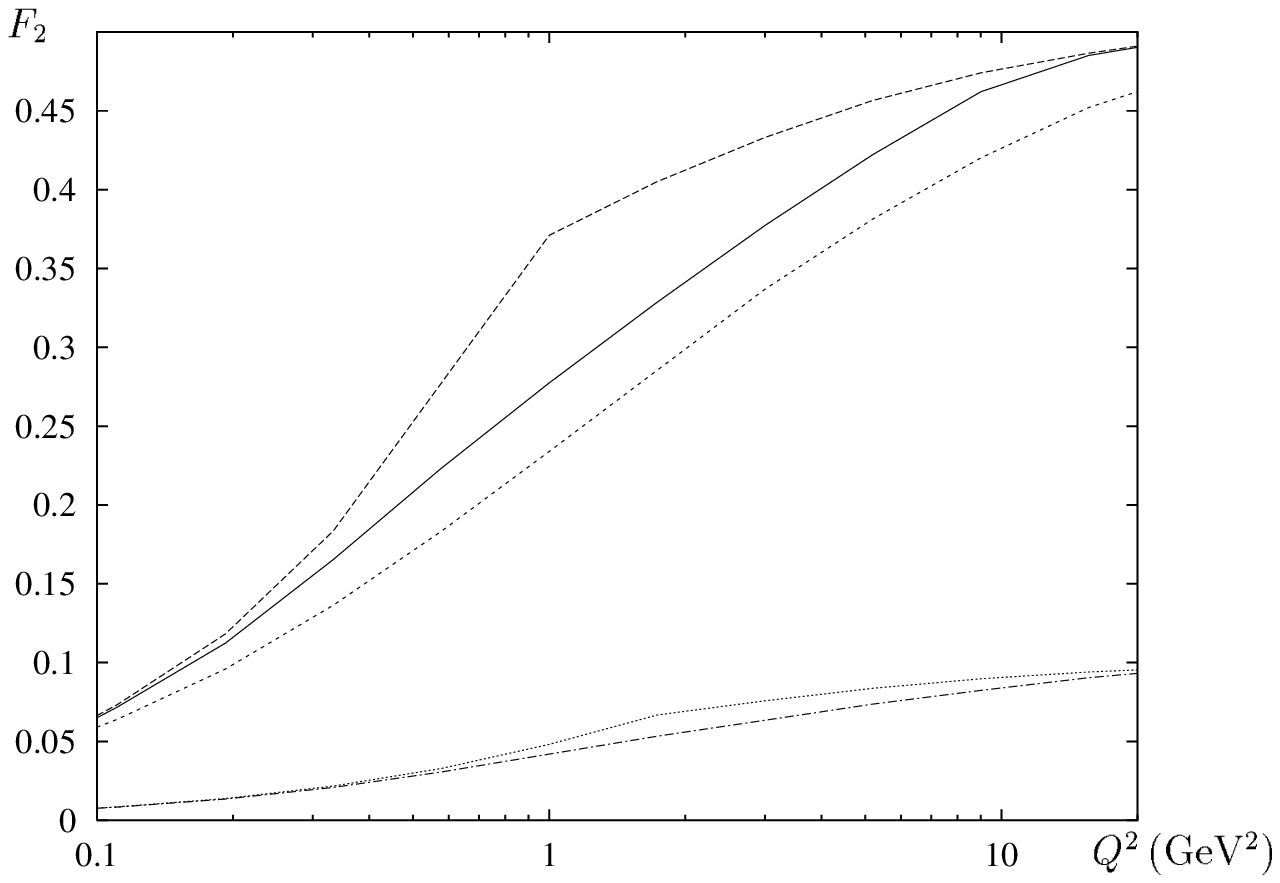}$$
$$\epsfxsize=13cm\epsfbox{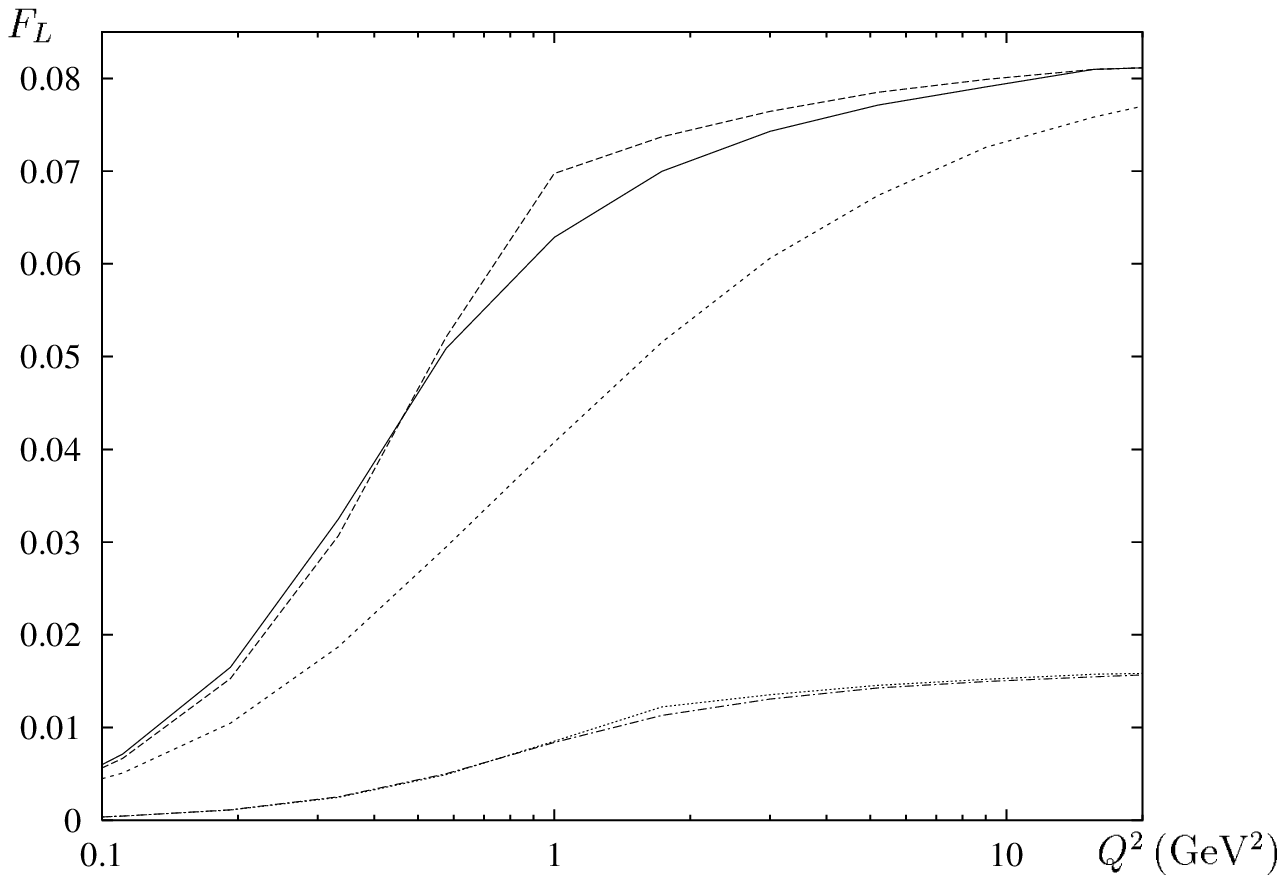}$$
\caption{(a) Theoretical calculation of $F_2$ as a function of $Q^2$.
The long-dash curve is the light quark contribution with the effective 
quark mass given in Eq.~(\protect\ref{effective-mass}). The solid 
curve is with the effective quark mass given in 
Eq.~(\protect\ref{z-dependent-mass}). The short-dash curve is with a 
fixed constituent quark mass $m_0=0.22$~GeV. The strange quark 
contribution with the effective mass given by 
Eq.~(\protect\ref{effective-s-mass}) is shown as a dashed dotted curve.
The dotted curve is with the effective strange mass of 
Eq.~(\protect\ref{z-dependent-s-mass}).
(b) Same study for the longitudinal structure function $F_L$.} 
\label{structure-function}
\end{figure}

In Fig.~\ref{structure-function}, we show the theoretical results
for $F_2^{u+d}(x_B=Q^2/W^2,Q^2)$ and $F_2^{s}(x_B=Q^2/W^2,Q^2)$ 
at a fixed energy $W=20$~GeV. At this energy and in the $Q^2$ range 
considered, the Reggeon contribution to $F_2$ is less than 
5\%~\cite{don94}.

The calculation in Fig.~\ref{structure-function} with fixed light 
quark mass $m_0$ shows a dependence similar to the logarithmic 
dependence $\ln(Q^2/m_0^2)$ expected for the $r^2$ dipole-proton 
cross section. This agreement is also quantitatively good. The 
sliding quark mass $m=m_{\rm eff}(Q^2)$ produces a steeper variation 
of $F_2$ for $0.1$~GeV$^2\le Q^2\le 1$~GeV$^2$, in agreement with the
variation of the mass in the logarithm. Above $Q^2=1.05$~GeV$^2$ 
where its value has gone to zero, the finiteness of $F_2$ is 
connected with the long distance behavior of the dipole-nucleon 
cross section. In this region of $Q^2$ the $z$ averaged integrand 
in Eq.~(\ref{T-integrand}) has a maximum in $r$ in the region 
$r=0.2$--$0.8$~fm and it extends to very large values of $r$, being only
damped because of the change of the quadratic behavior of $J_p=Cr^2$ at
small distances to a  power $1.5$--$1.8$. The maxima of the integrands
for the transverse and longitudinal cross sections in $r$ are not so
different but the profile is much faster decreasing in the longitudinal
case. In the case of a dependence of $m_{\rm eff}$ on 
$Q^2_{\rm eff}=4z(1-z)Q^2$ the resulting structure function $F_2$ in
Fig.~\ref{structure-function} interpolates smoothly the case of
$m_{\rm eff}(Q^2)$ between the minimal and maximal values of $Q^2$. It
lies about 10--20\% above the curve with constant $m_0$ for
$Q^2\ge0.2$~GeV$^2$. The strange quark structure function reaches at the
maximum $Q^2$ the asymptotic rate of 20\% of the light quark structure
function. The difference between the contribution given by the effective
mass of Eq.~(\ref{effective-s-mass}) and the one of
Eq.~(\ref{z-dependent-s-mass}) is qualitatively the same as in the light
quark case.

The longitudinal scattering is suppressed compared with the transverse
scattering in agreement with the $r^2$ estimates. Here the difference
between the $z$ dependent effective quark mass and the only $Q^2$
dependent quark mass is very small,  since the longitudinal photon wave
function suppresses the endpoints in the $z$-integration. Conversely the
difference with the fixed $m_0$ mass reaches about 30\% at
$Q^2=1$~GeV$^2$ and is more visible than in $F_2$.

\begin{figure}
$$\epsfxsize=13cm\epsfbox{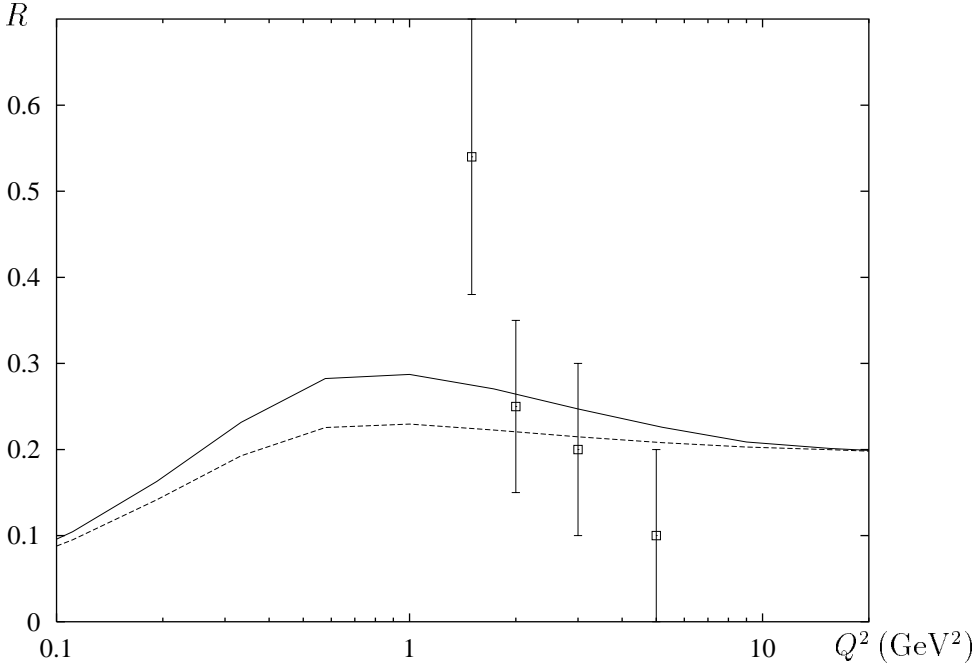}$$
\caption{$R=\sigma_L/\sigma_T$ as a function of $Q^2$. 
The solid curve is our expectation with $z$ dependent quark masses 
for light and strange quarks, Eq.~(\protect\ref{z-dependent-mass}) 
and Eq.~(\protect\ref{z-dependent-s-mass}), respectively.
The dashed curve is the result with the effective quark masses 
given in Eq.~(\protect\ref{effective-mass}) and  
Eq.~(\protect\ref{effective-s-mass}). Data are from 
NMC~\protect\cite{nmc}.}
\label{ratio}
\end{figure}

At the photoproduction point the ratio $R=\sigma_L/\sigma_T$ vanishes,
it then increases until $Q^2=1$~GeV$^2$. For the quark mass depending on
the virtuality $Q^2$ its behavior is more flat than when the quark mass
depends on $Q_{\rm eff}^2$. A computation with a fixed mass, $m_0$, and
the short distance behavior $J_p\propto r^2$ leads to a ratio similar in
shape with a maximum in $Q^2$ around $40\,m_0^2$. This result was
already obtained in the classic paper of Bjorken, Kogut and Soper on the
electroproduction of lepton pairs in a slowly varying external
field~\cite{bjo71}. Note that in this case a constant quark mass
$m_0=0.22$~GeV would give a maximum in $R=\sigma_L/\sigma_T$ at
$Q^2\approx 2$~GeV$^2$. The precise behavior of the ratio 
$R=\sigma_L/\sigma_T$ gives another signal for the chiral transition in
the experimental deep inelastic scattering data. We show in
Fig.~\ref{ratio} the ratio we get in our computation, combining light
and strange contributions. We compare this ratio to NMC results.
Unfortunatly experiments do not reach the small $Q^2$ transition region.

\begin{figure}
$$\epsfxsize=13cm\epsfbox{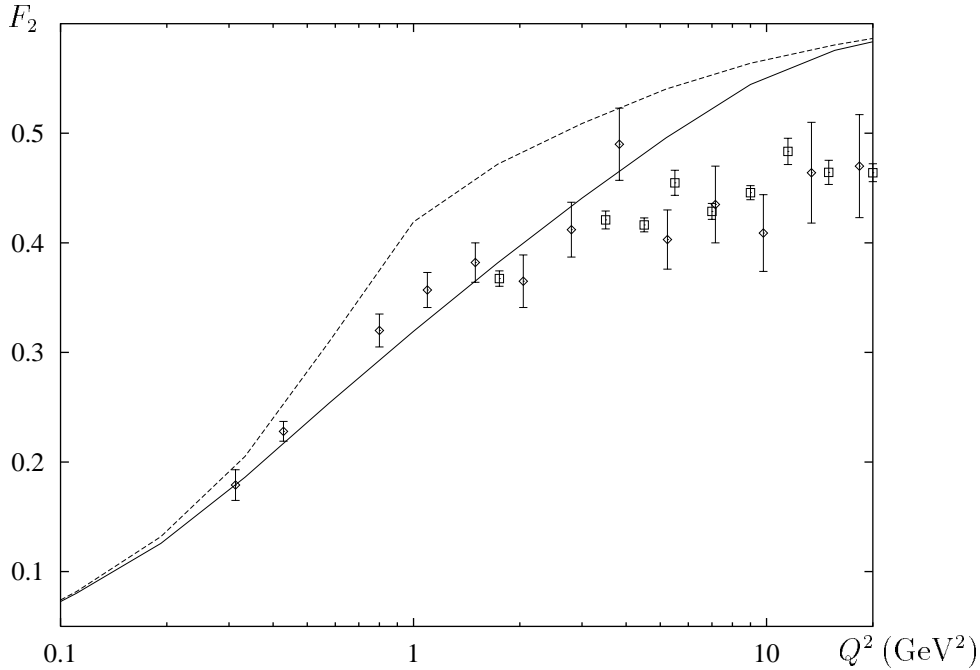}$$
\caption{Contribution of $u$, $d$ and $s$ to $F_2$ as a function of 
$Q^2$. Curves are as in Fig.~\protect\ref{ratio}. Squares are 
NMC results~\protect\cite{nmc} and diamonds are 
E665~\protect\cite{e665}.} 
\label{f2}
\end{figure}

We next combine the light and strange contribution to form the 
structure function $F_2=F_2^{u+d}+F_2^{s}$ as a function of 
$Q^2$ and for fixed energy $W=20$~GeV. In Fig.~\ref{f2}, we 
compare our results with data from both E665 and NMC at energies 
$18.5$~GeV$\le W\le 21.5$~GeV. The transition region is correctly 
described and the scheme with the $z$ dependent effective mass seems 
to be preferred. At large $Q^2$ region, our result overshoot data 
by 10--20\%. As we now show finite energy corrections may reduce cross
sections in this large $Q^2$ region.

\section{Modification at finite energy}\label{sec:fe}

\subsection{Slow quarks}

\begin{figure}[ht]
$$\epsfxsize=11cm\epsfbox{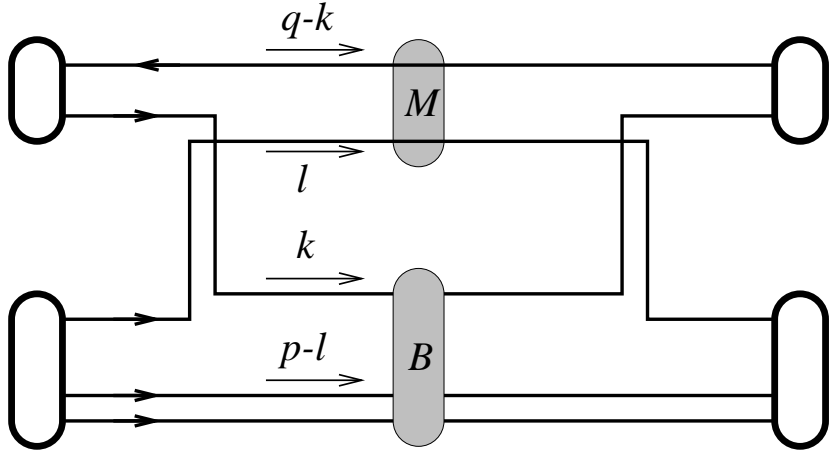}$$
\caption{The intermediate rearrangement for light quarks is determined
by the momenta of the collison partners as calculated in 
Eq.~(\protect\ref {threshold}).}
\label{intermediate-state}
\end{figure}

In the forward Compton amplitude, we integrate over configurations with
quark lightcone fractions varying from 0 to 1. For the transverse cross
section, there is a large contribution from aligned jet configurations,
where one quark carries most of the momentum and the other one a minute
fraction. At large but {\em finite} energy, however, the slow quark may
not carry enough energy to generate a {\em hadronic} final state.
Formally, the photon-proton total cross section is related to the
imaginary part of the forward elastic amplitude via the optical
theorem. In this amplitude we must sum only over the accessible
channels, i.e., we have to take into account energy conservation at
finite energies. To be more precise we  have to care about how the
energy is distributed in the physical color neutral final states. We
require that the intermediate quark-antiquark and quark-diquark states
(see Fig.~\ref{intermediate-state}) both have an invariant mass bigger
than a typical mesonic or baryonic state with mass 
$M_M=(M_{\rho},M_{K^*})$ or $M_B=(M_N,M_{\Lambda})$ respectively.
\begin{eqnarray}\label{threshold}
(k+p-l)^2&\approx&z_1(1-z_2)W^2\ge M_B^2\\
(q-k+l)^2&\approx& (1-z_1)z_2W^2\ge M_M^2.\nonumber
\end{eqnarray}

For the present discussion, we restore the internal degrees of
freedom of the proton, i.e., the cross section envisaged here is written
as in Ref.~\cite{dos97} with the full dependence on transverse and light
cone coordinates of the quarks in both the photon and the proton.
\begin{equation}\label{cross-section}
\sigma_{\gamma^* p}=2\int d^2{\bf b}
\int {dz_1d^2{\bf r}_1\over 4\pi}|\psi_\gamma(z_1,{\bf r}_1)|^2
\int {dz_2d^2{\bf r}_2\over 4\pi}|\psi_p(z_2,{\bf r}_2)|^2\,
J({\bf x}_1,{\bf x}_{\bar 1},{\bf x}_2,{\bf x}_{\bar 2})
\Theta(z_1,z_2,W),
\end{equation}
where 1 and 2 refer to the photon and nucleon sides, respectively. The
proton is considered in a simple quark-diquark picture which has proved
to be a good approximation in the framework of the stochastic vacuum
model. In Eq.~(\ref{cross-section}), we have inserted the threshold
factor $\Theta$ which realizes the requirement of Eq.~(\ref{threshold}):
\begin{equation}\label{threshold-factor}
\Theta(z_1,z_2,W)=\theta\Big[z_1(1-z_2)W^2-M_B^2\Big]\,
\theta\Big[(1-z_1)z_2W^2-M_M^2\Big].
\end{equation}

In the integrand of Eq.~(\ref{L-T-amplitude}), the effect of the
threshold factor is to generate a $z\equiv z_1$ dependent phase space
factor: 
$$
\Phi(z)=\int_0^1 dz_2\,f(z_2)\,
\theta\left[1-{M_1^2\over zW^2}-z_2\right]\,
\theta\left[z_2-{M_2^2\over(1-z)W^2}\right].
$$
In the integral, $f(z_2)$ represents the $z_2$ dependence as it results 
for the various term in Eq.~(\ref{cross-section}). Indeed, since the
$z_2$ dependence of the quantity $J$ is rather weak, $f(z_2)$ turns out 
to be given essentially by the square of the proton wave function, for
which we use $f(z_2)=252z_2^2(1-z_2)^6$. For large energies, 
$W\gg M_M+M_B$, $\Phi(z)=1$ for intermediate $z$ and rapidly decreases
to 0 when $z$ approaches the boundaries which at large enough $W$ read
$$
{M_B^2\over W^2}\le z\le 1-{M_M^2\over W^2}.
$$
For $M_B=M_p$, $M_M=M_\rho$ and $W=20$~GeV, the boundaries are
approximately $0.0022\le z\le 0.9985$. 

\begin{figure}[ht]
$$\epsfxsize=10cm\epsfbox{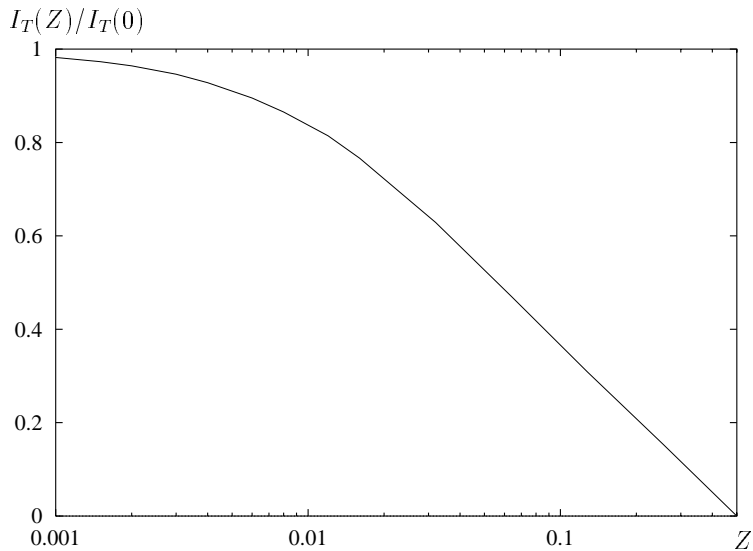}$$
\caption{
Interaction amplitude, $I_T(Z)/I_T(0)$, at $Q^2=3$~GeV$^2$ as a function
of the cut off $Z$ limiting the momentum fraction of quarks in the
photon. The effective mass used in this computation is the one given in
Eq.~(\protect\ref{z-dependent-mass}).}
\label{z-dependence}
\end{figure}

The threshold factor is therefore important in case where the endpoint
contribution is sizeable. Usual hadron wave functions suppress this
region in both hadron-hadron collisions and vector meson production,
thus rendering this effect unimportant at large energy, 
$s\ge 100$~GeV$^2$. In inclusive photon-hadron scattering, however, this
endpoint region cannot be overlooked. The importance of the various
region in $z$, in the full integral Eq.~(\ref{L-T-amplitude}), may be
studied by varying the lower limit $Z$ of the integration over quark
light cone momenta in the photon wave function. Recall that the
threshold condition for light quarks gives a lower limit $z\ge 0.0022$ 
for $W=20$ GeV as calculated above and that this lower limit goes like
$1/W^2$. Since the integrand is symmetric in $z$, the upper integration
limit can be restricted to $z\le 0.5$. The resulting function $I_T(Z)$ 
$$
I_T(Z)=\int_Z^{0.5} dz\int_0^{\infty}rdr{\cal I_T}(z,r),
$$
is shown in Fig.~\ref{z-dependence} for the transverse cross section. 

In order to understand what happens in the endpoint region, it is
instructive to study the behavior of the amplitude for a simplified
$J_p(z,r)$ behaving like $C\,r^n$ with $n=1$ or 2. Our actual $J_p$ is
in some sense interpolating between those two choices, the power $n=2$
corresponding to the short distance dipole behavior already mentioned in
Eq.~(\ref{dipole-behavior}). With such a simple dependence, one can
perform the $r$-integral analytically:
\begin{eqnarray}
\int rdr\,{\cal I}_L&=&K{4z^2(1-z)^2Q^2\over[z(1-z)Q^2+m^2]^{1+n/2}},\\
\label{T-approximated}
\int rdr\,{\cal I}_T&=&K{(1+2/n)[z^2+(1-z)^2]\over[z(1-z)Q^2+m^2]^{n/2}}
+K{m^2\over[z(1-z)Q^2+m^2]^{1+n/2}},
\end{eqnarray}
where the overall constant is 
$$
K={N_cC\over4\pi^2} 2^{n-1}{\Gamma^4(1+n/2)\over (n+1)!}.
$$
For $m^2\ll Q^2$, one can focus on the first term in
Eq.~(\ref{T-approximated}). At $Z\to 0$, the quantity $Q^n I_T(Z)$ 
behaves, for $n=2$, like
$$
Q^2 I_T(Z)\propto O(1)-\ln[Z+m^2/Q^2],
$$ 
and, for $n=1$, like 
$$
Q\,I_T(Z)\propto O(1)-\sqrt{Z+m^2/Q^2}.
$$
In Fig.~\ref{z-dependence}, we see the transition from the logarithmic
behavior given in case $n=2$ to a linear behavior at very small $Z$. The
limit $Z=0$ is however approached like in the case $n=1$: it does not
have the dramatic dependence on $m^2$ exhibited by the $n=2$ case. For
$Q^2\le 4m^2$, the integrands become more or less flat in $z$ and the
$Z$ behavior is linear in the whole range.

Because the photon size parameter, $\varepsilon^2=z(1-z)Q^2+m^2$, is the
important scale in $\cal I$, the behavior changes in the neighborhood of
$Z=m^2/Q^2$. The region where the threshold suppression in photon-proton
collisions is sizeable is therefore given by
$$
m^2/Q^2\lsim M^2/W^2,
$$ 
i.e.,
$$
m^2/M^2\lsim x_B.
$$
For $M=M_p$ and $W=10$~GeV, this shows that the effect becomes sizeable
when $Q^2\ge 0.5$~GeV$^2$, for an effective quark mass $\sim 0.1$~GeV.
If we were to consider a current quark mass below this value of $Q^2$,
the effect would show up at even a much smaller value of $Q^2$.

\subsection{Threshold effect in the cross section}

\begin{figure}
$$\epsfxsize=13cm\epsfbox{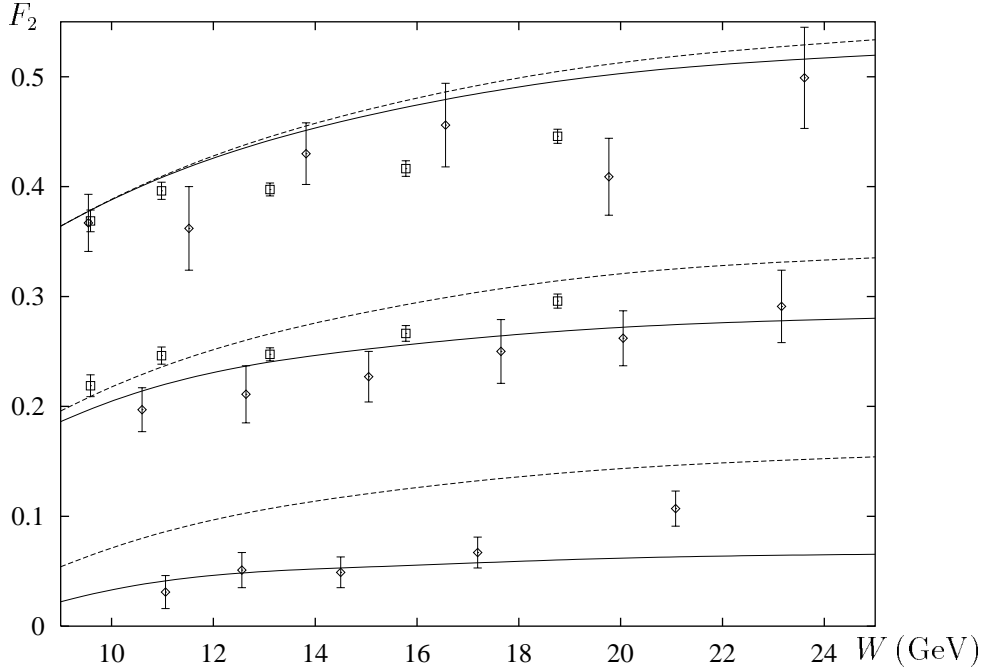}$$
\caption{Contribution of $u$, $d$ and $s$ to $F_2$ as a function of 
$W$, for $Q^2=$1, 3 and 9~GeV$^2$, from bottom to top ($0.25$ has been 
substracted to the first set and $0.15$ to the second). The solid lines
represent our result for $z$ dependent effective masses and the dashed
lines correspond to $Q^2$ dependent effective masses. Squares are 
NMC results~\protect\cite{nmc} and diamonds are 
E665~\protect\cite{e665}.} 
\label{f2-threshold}
\end{figure}

We now compare with data our computed cross section modified by the
threshold effects from slow quarks. For photoproduction, the change is
negligible. For electroproduction, it is best to consider fixed $Q^2$
and vary the energy $W$ to see the threshold effect. In
Fig.~\ref{f2-threshold} we show the variation of $F_2$ for 
9~GeV$\le W\le 25$~GeV separately for $Q^2=1$, 3 and 9~GeV$^2$. (For
convenience we shifted downward the first two sets by respectively 
$0.25$ and $0.15$.) We have checked that for these values of 
$Q^2$ the inclusion
of the Reggeon contribution given in Ref.~\cite{don94} only modifies
weakly the trend of our curves for $W>10$~GeV. In each case we present 
results combining $u$, $d$ and $s$ contributions for both schemes: $z$
dependent quark masses, Eqs.~(\ref{z-dependent-mass})
and~(\ref{z-dependent-s-mass}), and $z$ independent quark masses,
Eqs.~(\ref{effective-mass}) and~(\ref{effective-s-mass}). The effect is
relatively stronger for the latter scheme since the endpoint
contributions are clearly more important in this case. The difference
between the two schemes decreases at small $W$ where only intermediate
$z$ are taken into account. As anticipated the threshold effect is also
more important at large $Q^2$ and at $Q^2=9$~GeV$^2$ the suppression of
the cross section is typically 10\% at $W=20$~GeV and reaches 30\% at
$W=10$~GeV. At this high $Q^2$ the effective quark mass has practically
gone to zero in both schemes and the large distances are cut off by the
threshold condition. 

\begin{figure}
$$\epsfxsize=13cm\epsfbox{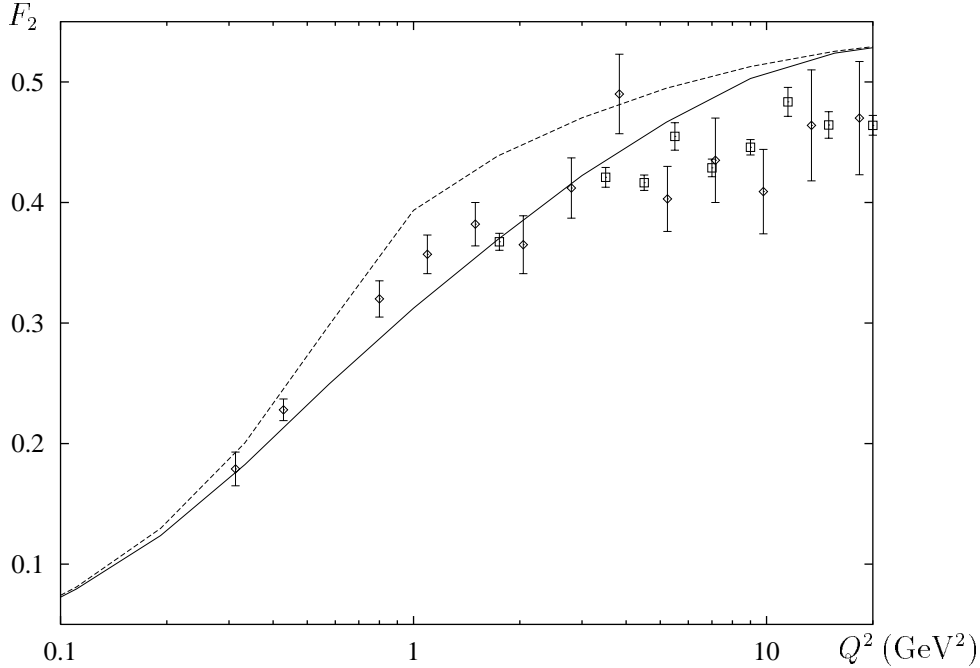}$$
\caption{Contribution of $u$, $d$ and $s$ to $F_2$ as a function of 
$Q^2$, for $W=$20~GeV. Curves and data are as in
Fig.~\protect\ref{f2}}
\label{f2-20}
\end{figure}

To complete our study we give in Fig.~\ref{f2-20} our result for 
$F_2(x_B=Q^2/W^2,Q^2)$ at $W=20$~GeV including the threshold effect. 
The plot may be compared with Fig.~\ref{f2} and we see that the
qualitative aspect of the transition from low to high $Q^2$ remains. In
general the large $Q^2$ range is sensitive to the threshold effect
whereas the small $Q^2$ range tests more the effective quark mass.

\section{Discussion and Conclusions}

We have computed the total photon proton cross section in a model of
nonperturbative QCD. For values of the virtual photon mass
$Q^2>2$~GeV$^2$ our input is the  treatment of two Wilson loops in
Minkowski space-time within a special model of nonperturbative QCD which
approximates the infrared behavior by a Gaussian stochastic process
determined by a non-local gauge invariant gluon field correlator. The
latter one is essentially given by the local gluon condensate and the
correlation length. In order to fix the size distribution of one loop a
proton valence quark wave function has to be introduced. In principle
all parameters of the model can be determined  by sources other than
high energy scattering, namely lattice gauge calculations and low energy
phenomenology. In practice the errors in the parameters still
necessitate some adjustment to high energy scattering data. In our case
we have chosen the determination of Ref.~\cite{dos97} based on the
proton-proton total cross section and the logarithmic slope of the
elastic $pp$ cross section at zero momentum transfer. Although the
parameters turn out to be rather stable there remains still a certain
variation in a range of a few percent for the correlation length and
20--30\% for the gluon condensate and the proton radius. The model gives
energy independent cross sections, contrary to the slight $s^{0.08}$
energy dependence seen in low $Q^2$ experiments. Since the parameters
are related to the ISR energies, $W\approx20$~GeV, we also compare the
photon cross sections calculated here with experimental values around
that energy. We have already used this procedure in order to calculate
diffractive electroproduction of vector mesons with very good results. 

In the present work we have extented the $Q^2$ range down to $Q^2=0$. In
order to do that we have constructed nonperturbative photon wave
functions essentially by introducing a virtuality dependent constituent
quark mass. We were encouraged to such a simple procedure by model
investigations of harmonic oscillator Green functions.
In our method we adjusted the value of the momentum dependent quark mass
to reproduce the phenomenological two point function of the vector
currents. The development of a quark mass at large distances has been
seen also in calculations with the instanton liquid model~\cite{shu89}.
We repeat that none of our input parameters was in any way related or
adjusted to electroproduction phenomenology. 

Our approach differs in several ways from other investigations. For the
treatment of the soft exchanges, Nikolaev and Zakharov have adopted a 
phenomenological two-gluon-exchange model~\cite{nik91}. Their treatment
also gives importance to the $q\bar{q}$ photon wave function and their
dipole-proton cross section has a $r^2$ behavior at short distance and
saturates in the $r=1$--2~fm region. To suppress the contribution from
large distance in the photon wave function they cut off the large size
component using a smooth gaussian $\exp (-r^2/R_c^2)$ with a confinement
size parameter $R_c\approx 1.5\,$fm. In Ref.~\cite{nik91}, a current
quark mass $m_{u,d}=10$~MeV was used, but in a later study of the same
authors on the BFKL pomeron~\cite{nik94} a constant light quark mass
$m_{u,d}=150$~MeV was considered. As we have seen in our study such a
value can help to limit the extent of the photon wave function, but at
large $Q^2$ there is no reason that the light quarks have masses
different from the current quark masses. In the region of transverse
distance probed, we have seen that the nonperturbative features of the
gluon correlators are important and we think that a perturbative
two-gluon exchange model cannot be trusted.

Concerning the transition to small $Q^2$, we have shown an approach
different from  vector meson dominance (VMD) frequently used in the low
$Q^2$ range. We claim  that with increasing $Q^2$ one would have to put
a rapidly growing number of resonances into the VMD model which thereby
becomes untractable. Our scheme of quark-hadron duality exploits the
knowledge about the residue and the mass of the lowest vector meson
state contributing to the vector current two-point function.

Based on the transition between a VMD description of $F_2$ at small
$Q^2$ and the partonic one at large $Q^2$ Badelek and Kwieci\'nski have
proposed to represent the proton structure function via dispersion
relation~\cite{bad92}. The construction adopted here shares similarities
with their approach although the latter makes no connexion to the notion
of wave function which we need for the microscopic description of
diffractive scatterings. In their approach the partonic contribution to
$F_2$ is extracted from structure function analysis at large $Q^2$ and
thus naturally fulfills perturbative QCD evolution at large virtualities
where this evolution is experimentally observed. Such perturbative
corrections are not implemented in our approach.

\begin{figure}
$$\epsfxsize=13cm\epsfbox{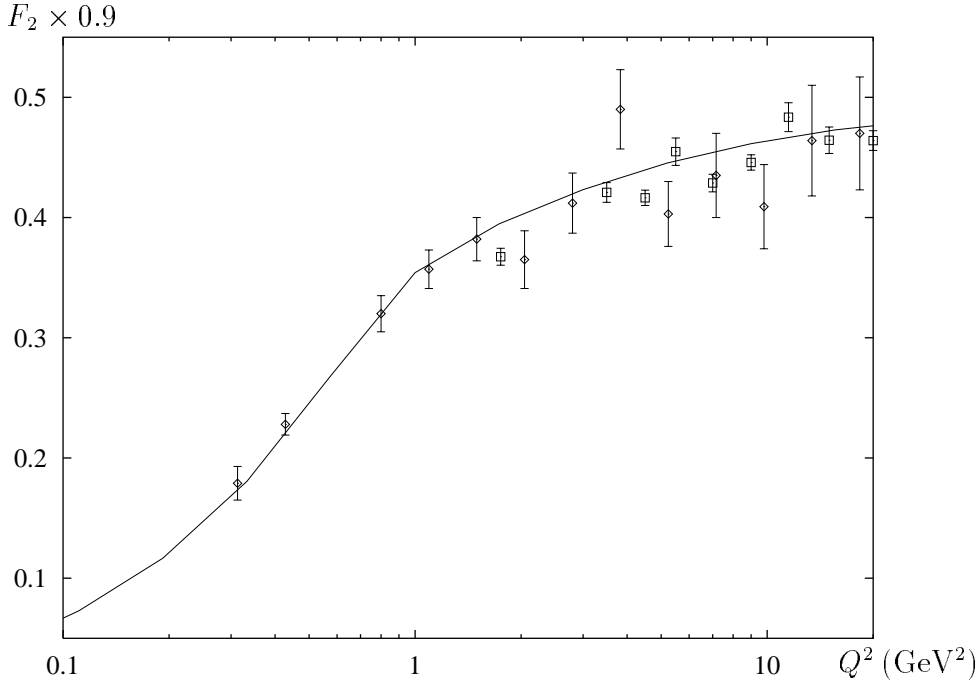}$$
\caption{Contribution of $u$, $d$ and $s$ to $F_2$ as a function of 
$Q^2$, for $W=20$~GeV. The curve is the result shown in 
Fig.~\protect\ref{f2-20} and for the $Q^2$ dependent quark masses of 
Eqs.~(\protect\ref{effective-mass}) and~(\protect\ref{effective-s-mass})
(dashed curve in Fig.~\protect\ref{f2-20}) multiplied by $0.9$. This
rescaling would correspond to a 5\% decrease of the gluon condensate.
Data are as in previous figures.} 
\label{f2-resc}
\end{figure}

Our results are encouraging. Without any adjustment they agree 
with experiments over the full $Q^2$ range from 0 to 20~GeV$^2$  within
10--20\%. For the high $Q^2$ range we have shown how finite energy
corrections may account for at least a part of the discrepancy without
changing any of the model parameter and without affecting the good
description of the transition region where the modification of the
perturbative $q\bar{q}$ wave function is at work. It is rather difficult
to decide with present data what is the best scheme in our approach,
i.e., whether or not the effective mass depends on the quark light cone
fraction. In order to put our result in a larger perspective, we show in
Fig.~\ref{f2-resc} the curve one deduces from Fig.~\ref{f2-20} with the
z~independent masses, Eqs.~(\ref{effective-mass})
and~(\ref{effective-s-mass}), if we adjust the normalization by $-$10\%
(which would amount, e.g., to a 5\% reduction of the gluon
condensate). A rescaling of the curve in Fig.~\ref{f2} by $-$15\% shows
the same pattern. We describe almost perfectly the $Q^2$ dependence of
the data in the whole $Q^2$ range examined. This points toward the
possibility that the chiral transition is already seen in present data,
the kink in the data at $Q^2=1$~GeV$^2$ being related to the vanishing
of the quark mass at that value of $Q^2$ (chiral restoration). The
photon-proton cross section could also be easily adjusted by a slight
decrease of the constituent quark mass. 

Because of the fundamental importance of chiral symmetry breaking for
hadron physics, the question of chiral symmetry restoration with
increasing virtuality $Q^2$ of the photon deserves to be studied in more
detail. More systematic data for $F_2(x_B,Q^2)$ and $F_L(x_B,Q^2)$ at
fixed $W$ and varying $Q^2$ would help in deciding whether a marked
change occurs between the low momentum domain and the perturbative
domain in inelastic electron scattering. It may be that chiral symmetry
restoration is seen more easily in electroproduction at a given
virtuality $Q_0^2$ than in heavy ion collisions at a finite temperature
$2\pi T=Q_0$.

\bigskip

We thank B. Kopeliovich, G.Kulzinger, O. Nachtmann and M. Rueter for
critical discussions.
T. G. was carrying out his work as part of a training Project of the
European Community under Contract No. ERBFMBICT950411.
This work has been partially funded through the European TMR Contract
No.~FMRX-CT96-0008: Hadronic Physics with High Energy Electromagnetic
Probes, and through DOE grant No.~85ER40214.

\end{document}